\documentclass[twocolumn,times]{aastex62}
\usepackage{amsmath}
\usepackage{amssymb}

\usepackage{bm}
\allowdisplaybreaks[4]

\def\be {\begin{equation}}
\def\ee {\end{equation}}

\def\tt {\{t\}}

\usepackage{color}
\usepackage{booktabs}

\begin{document}

\title{\large{\textbf{Giant impact between high-viscosity Theia and low-viscosity proto-Earth: Origin of lunar isotopic crisis}}}

\correspondingauthor{Wenshuai Liu}
\email{674602871@qq.com}

\author{Wenshuai Liu}
\affiliation{School of Physics, Henan Normal University, Xinxiang 453007, China}



\begin{abstract}
According to the giant impact theory, the Moon was formed by accretion of the debris disk that resulted from the collision between Theia and the proto-Earth. Although this theory accounts for most characteristics of the Earth-Moon system, numerical simulations of impacts between a planetary embryo and the accreting proto-Earth indicate that more than 40 percent of the material in the circum-terrestrial disk generated by such an impact originates from the impactor. This poses a challenge for the giant impact theory in explaining the Moon's Earth-like isotopic composition, a discrepancy known as the lunar isotopic crisis. Since terrestrial planets were melted one or more times during accretionary processes, magma ocean on the surface of a growing planet would appear. Small terrestrial planets with magma ocean cool faster than large ones, resulting that the viscosity of small terrestrial planets is larger than that of large terrestrial planets still covered by magma ocean. Here, it shows that giant impact between a high-viscosity Theia and a low-viscosity proto-Earth could produce a circum-terrestrial debris disk predominantly composed of material from the proto-Earth without violating the angular momentum constraint of modern Earth-Moon system. The theory proposed here may provide a natural way of explaining the lunar isotopic crisis.
\end{abstract}



\section{Introduction}

According to the giant impact theory \citep{1,2}, the Moon formed from material accreted from a circum-terrestrial debris disk, which was generated after a planet-sized body collided with the proto-Earth. Numerical simulations based on this theory have shown that this disk from which the Moon originated contains a substantial fraction of material derived from the impactor \citep{15,16,17}. Since different bodies in the Solar System have distinct compositions, the canonical giant impact model \citep{15} would predict a lunar composition significantly different from that of Earth. However, isotopic comparisons between Earth and Moon samples reveal a remarkable similarity \citep{18,19,20,21,22,23}, a discrepancy known as the lunar isotopic crisis. To explain this isotopic similarity, either the Moon-forming material must have equilibrated with, or originated from, Earth's mantle after the impact \citep{24,22}, or the impactor itself must have had an isotopic composition identical to that of Earth \citep{26}. Nonetheless, none of these models have fully accounted for all geochemical observations, highlighting the need for further research into the complex processes that shaped the Earth-Moon system.

\cite{27} introduced a variant of the giant impact hypothesis involving a rapidly rotating proto-Earth. In this scenario, a larger portion of proto-Earth material can be transferred to the circum-terrestrial debris disk, which becomes dominated by proto-Earth matter. This disk is massive enough to potentially form the Moon following a collision between a fast spinning proto-Earth and a body slightly smaller than Mars. While this idea has the potential to explain the lunar isotopic composition, it yields an angular momentum exceeding that of the present Earth-Moon system. Other high-energy, high-angular momentum models, such as the Synestia model \citep{28,29} and the collision of two comparably massive bodies \citep{30} also produce excessive angular momentum, which would need to be reduced by some feasible mechanism, still a matter of debate. The high-energy, high-angular momentum model \citep{27} is designed to produce a circum-terrestrial debris disk primarily composed of proto-Earth mantle material, thereby explaining the Earth-like isotopic composition of the Moon.

During their formation and early evolution, rocky planets may experience several episodes of extensive melting. The resulting magma oceans are generated by the radioactive decay of short-lived isotopes, the release of gravitational energy accompanying core formation, and large impact events. Small planets cool faster than larger ones due to small planets' larger surface area to volume ratio. Therefore, solidification of magma ocean of small planet would be quicker than that of larger ones, resulting that small planet may already solidify while magma ocean are still on larger planets, and that the viscosity ($\mathrm{10^{21} Pa*s}$) of the mantle of the small planets would be far larger than that ($\mathrm{0.1 Pa*s}$ to $\mathrm{10 Pa*s}$) of larger ones which may still be covered by magma ocean.

\section{Viscous Giant impact}

The numerical approach for investigating giant impacts is smoothed particle hydrodynamics (SPH). In such simulations, both the target and the impactor are represented by particles whose dynamics are governed primarily by gravitational forces and material pressure. The impact scenario shown in Figure 1 is specified by the impact parameter $b=\sin(\beta)$ and the speed at first contact of the impactor with the target's surface $v_c$. We set the initial position of the impactor in order to let the contact occur 2400 seconds after the simulation's start. $\beta$ is set to be $45^o$ and $v_c$ to be $v_{esc}$ where $v_{esc}=\sqrt{2G(M_t+M_i)/(R_t+R_i)}$ is the mutual escape speed of the system. Here, we set $M_t=M_E$ and $M_i=0.1M_E$ where $M_E$ is the mass of the Earth.

\begin{figure}
            \includegraphics[width=0.5\textwidth]{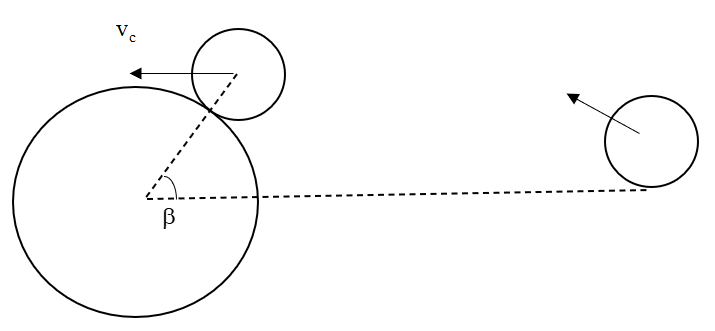}
\caption{The big circle represents the target planet with mass $M_t$ and the small circle is the impactor planet with mass $M_i$. $v_c$ is the speed at first contact of the impactor with the target's surface. The target planet rotates clockwise.}
\label{fig:figure1}
\end{figure}

To determine the radii of the non-rotating target and impactor, each body is divided into a mantle and an iron core, accounting for $70\%$ and $30\%$ of the mass, respectively. In reality, the outer portion of the proto-Earth may have existed as a magma ocean due to planetesimal accretion and multiple giant impacts. Here, we adopt Tillotson equations of state \citep{151} for the granite mantle and the iron core. We use OpenSPH \citep{900} with Barnes-Hut algorithm \citep{901} to generate the initial non-rotating target and impactor profile with about $10^5$ SPH particles.

We use miluphcuda \citep{150} to simulate the giant impact. The artificial viscosity is taken to be 1.5, the physical viscosities of Theia and proto-Earth are set to be $\mathrm{10^{14}Pa*s}$ and $\mathrm{10Pa*s}$ in the modified miluphcuda, respectively. For comparison, the physical viscosities of Theia and proto-Earth are both set to be $\mathrm{10Pa*s}$ in order to show the effect of the physical viscosity contrast on the giant impact.

The simulation with such high physical viscosity is so time consuming that the simulation of giant impact is terminated after 1.5 hours in order to preliminarily illustrate the effect of physical viscosity contrast on the composition of the debris ejected by the giant impact. The results are shown in Figure 2. It shows from Figure 2 that giant impact between a high-viscosity Theia and a low-viscosity proto-Earth could significantly affect the composition of debris where about $95\%$ materials originate from proto-Earth. Although the high viscosity of Theia is taken to be $\mathrm{10^{14}Pa*s}$ here, the composition of the debris disk will be totally made of materials of proto-Earth if Theia's viscosity is taken to be extremely large value of $\mathrm{10^{21}Pa*s}$. It's like throwing a stone into water, the stuff that splashes out is all just water.

In order to show a complete process of viscous giant impact (VGI) with fast simulation, we use miluphcuda \citep{150} to simulate a VGI of a $~1/10$ Moon-mass body with a Moon-mass target. $\beta$ is set to be $45^o$ and $v_c$ to be $1.4v_{esc}$. Each body is divided into a mantle and an iron core, accounting for $70\%$ and $30\%$ of the mass, respectively. Tillotson equation of state \citep{151} is adopted for the granite mantle and the iron core. The physical viscosities of the core of the $~1/10$ Moon-mass body and Moon-mass target are set to be $\mathrm{10^{13}Pa*s}$ while the physical viscosities of the mantle of the $~1/10$ Moon-mass body and Moon-mass target are set to be $\mathrm{10^{13}Pa*s}$ and $\mathrm{10Pa*s}$. The results are shown in Figure 3 where the debris disk is made of the Moon-mass target's mantle.

\begin{figure*}
   \begin{center}
     \begin{tabular}{cc}
            \includegraphics[width=0.45\textwidth]{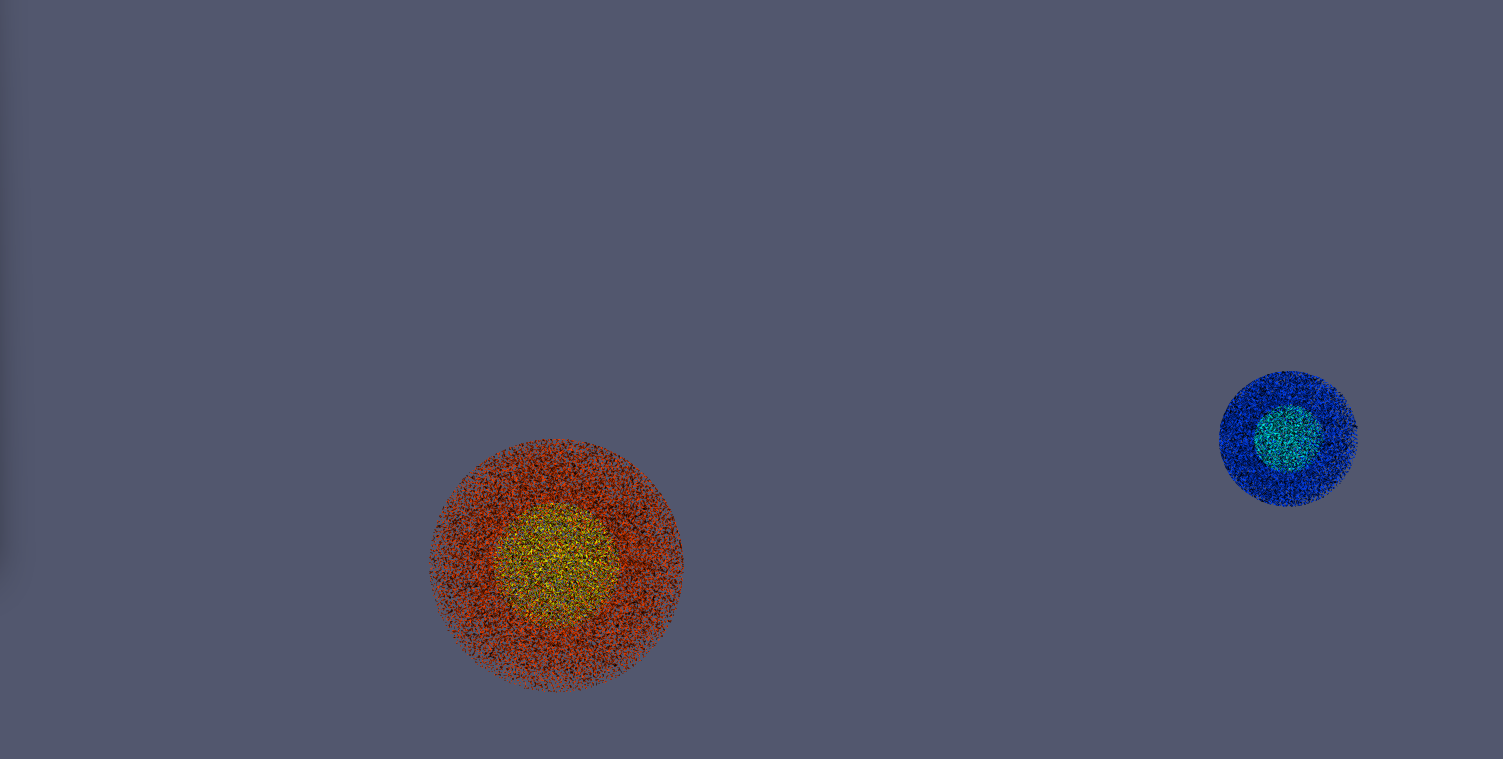}
            \includegraphics[width=0.45\textwidth]{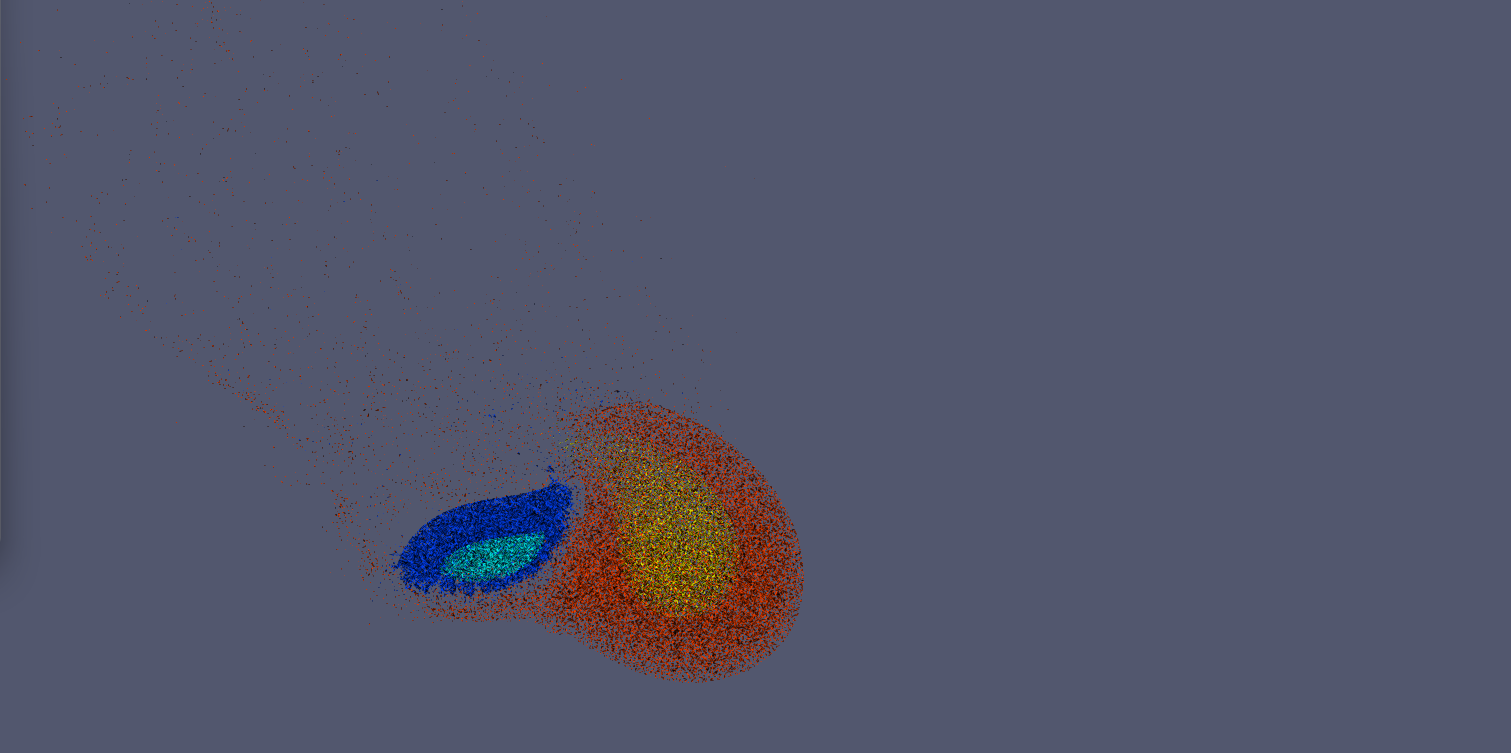}\\
            \includegraphics[width=0.45\textwidth]{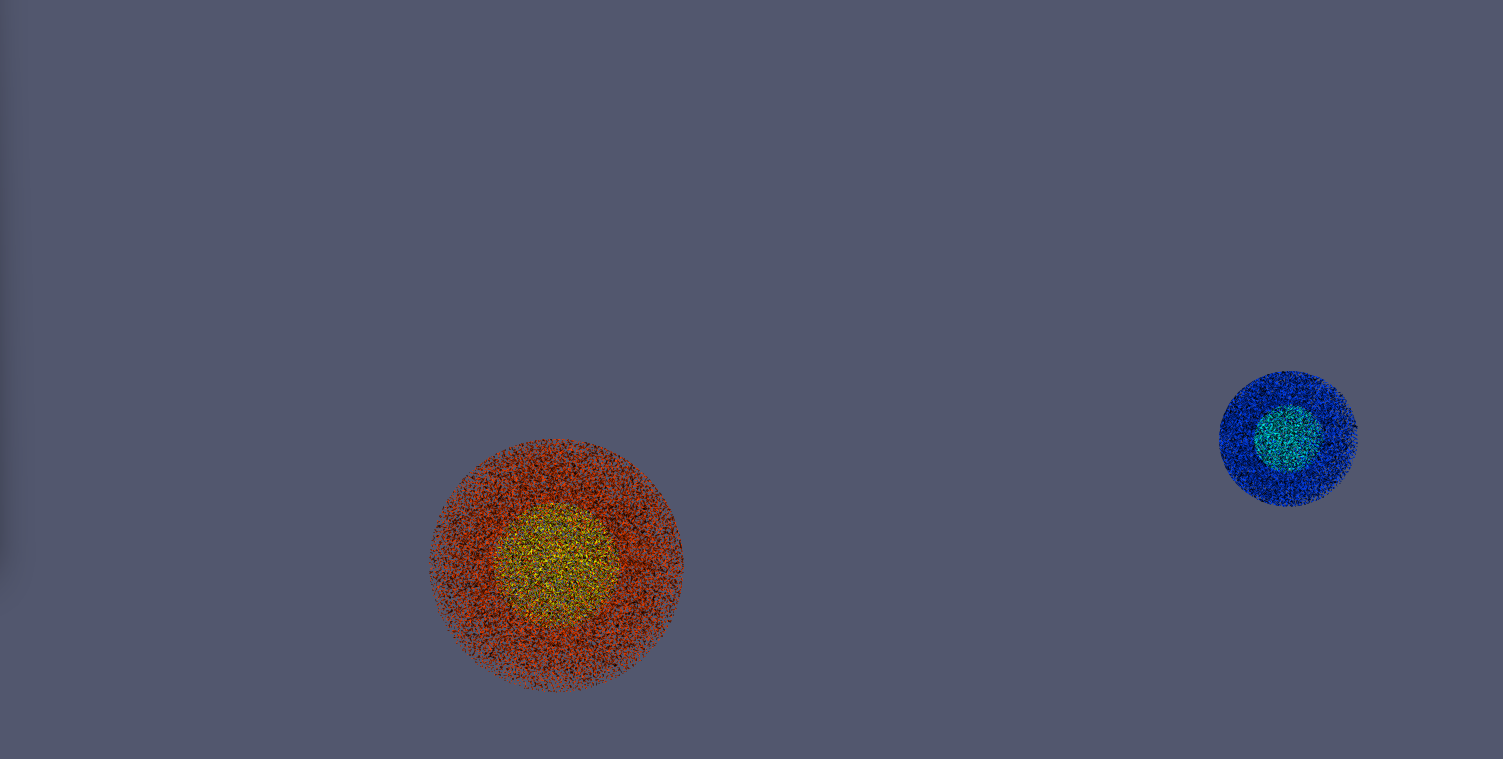}
            \includegraphics[width=0.45\textwidth]{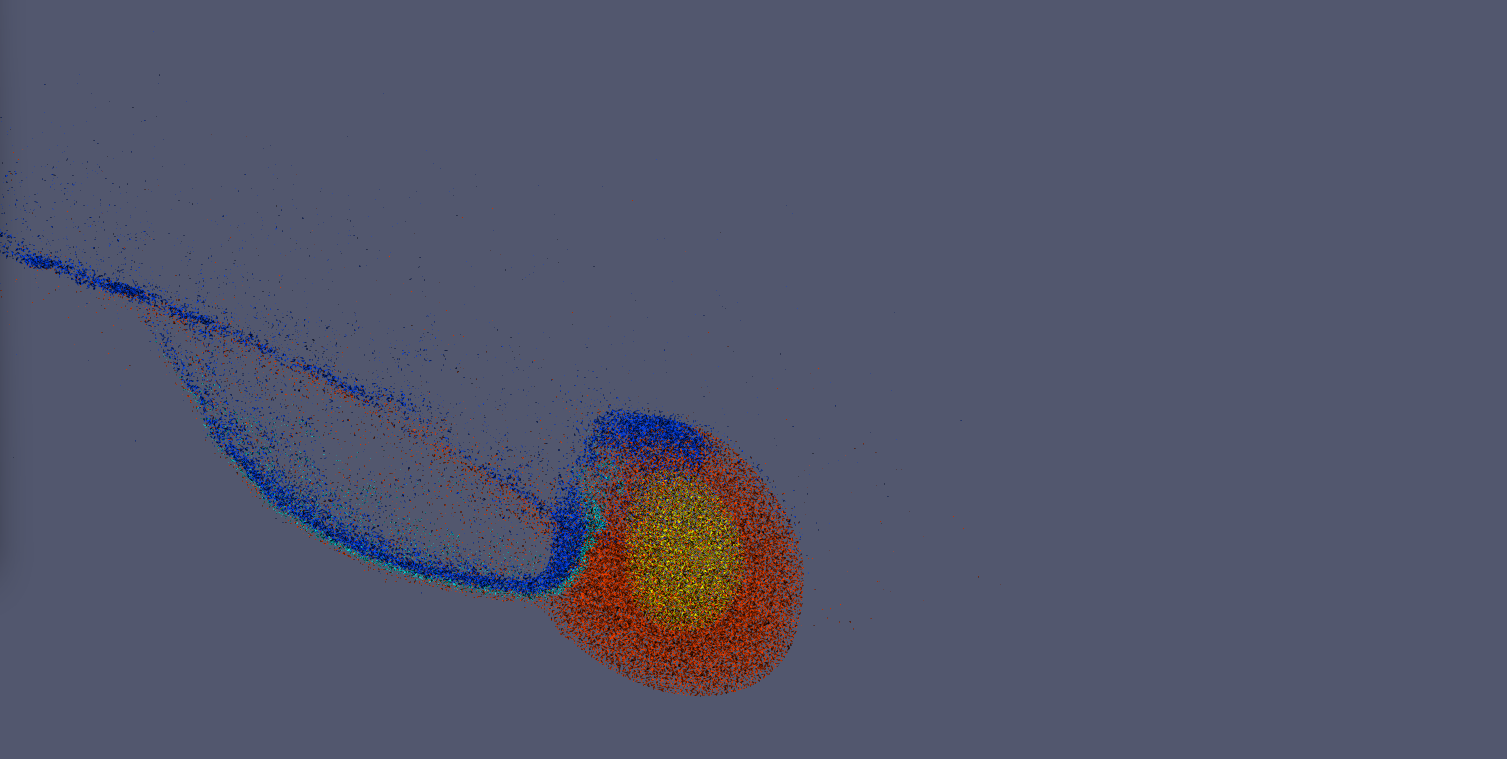}
            \end{tabular}
   \end{center}
\caption{Up and bottom panel represent the giant impact with viscosity contrast and same viscosity, respectively. Left and right show the results with simulation time of 0 hr and 1.5 hr, respectively.}
\label{fig:figure3}
\end{figure*}

\begin{figure*}
   \begin{center}
     \begin{tabular}{cc}
            \includegraphics[width=0.5\textwidth]{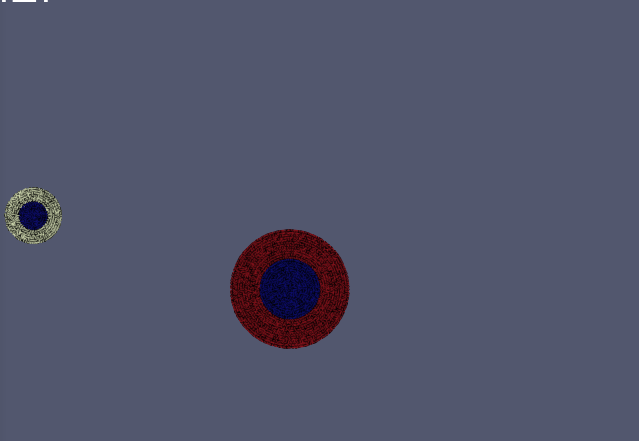}
            \includegraphics[width=0.5\textwidth]{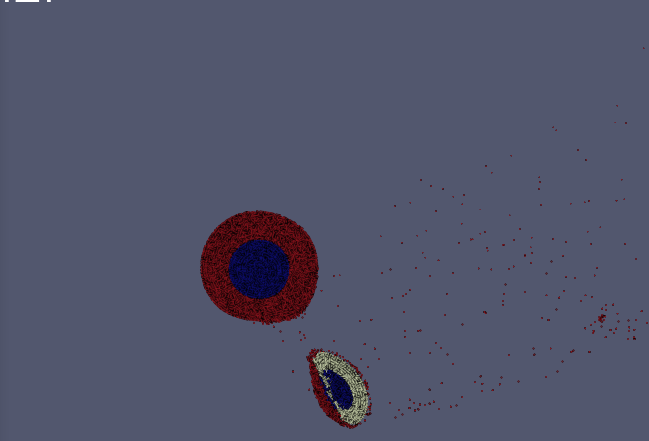}\\
            \includegraphics[width=0.5\textwidth]{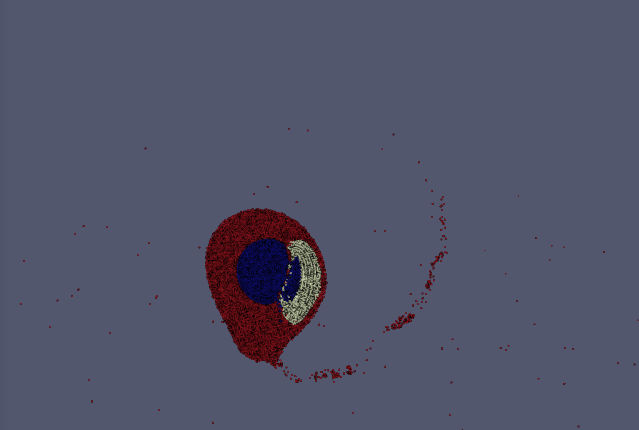}
            \includegraphics[width=0.5\textwidth]{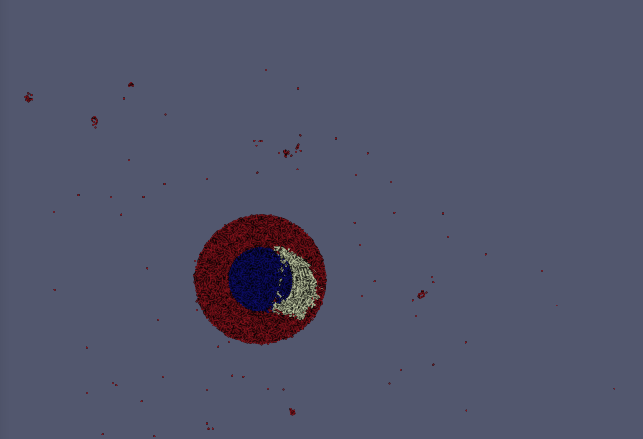}
            \end{tabular}
   \end{center}
\caption{Blue particles are the core's materials, yellow particles and red particles are the materials of the mantle of the $~1/10$ Moon-mass body and Moon-mass target, respectively. Up left, up right, bottom left and bottom right show the results with simulation time of 0 hr, 2.83 hr, 6.33 hr and 16.67 hr, respectively.}
\label{fig:figure3}
\end{figure*}

\section{Discussions}
We propose a new theory for the origin of the Moon's Earth-like isotopic composition in this work. Giant impact between a high-viscosity Theia and a low-viscosity proto-Earth could significantly affect the composition of debris disk. The larger the viscosity contrast, the higher portion of materials of proto-Earth in the debris disk. It can be deduced that the composition of the debris disk produced by giant impact will be completely made of materials of proto-Earth if Theia's viscosity is extremely large ($\mathrm{10^{21}Pa*s}$) and proto-Earth is covered by deep magma ocean. Furthermore, since Theia would reach solid state, material strength would enhance the result presented in this work. Due to the large viscosity of Theia, the complete VGI simulations with high viscosity Theia with material strength and low viscosity proto-Earth will be conducted very soon using miluphcuda. Besides giant impact by high viscosity Theia, low viscosity proto-Earth may experience multiple impact by high viscosity Moon-mass to Mars-mass planets.

Given that smaller bodies such as Vesta and Mars typically have mantles with a higher iron content than Earth's mantle \citep{27,31}, it is plausible that Theia possessed a mantle enriched in iron relative to that of the proto-Earth \citep{32}, meaning that the density of materials of Theia's mantle is greater than that of proto-Earth's mantle and that material originating from Theia's mantle would sink to the lower layer of Earth after giant impact.


\end{document}